\documentclass[twocolumn,prd,epsf]{revtex4-1}
\usepackage{amsfonts}
\usepackage{amsmath}
\usepackage{amssymb,amsbsy}
\usepackage{bm}
\usepackage{makeidx}
\usepackage{latexsym}
\usepackage{graphicx}
\usepackage{epsfig}
\usepackage{subfigure}
\usepackage{dsfont}
\usepackage{mathcomp}
\usepackage{color}
\usepackage{mathrsfs}

\begin{document}

\title{Neutron Stars as Dark Matter Probes.}%

\author{Arnaud de Lavallaz}
\email{arnaud.de\_lavallaz@kcl.ac.uk}
\author{Malcolm Fairbairn}
\email{malcolm.fairbairn@kcl.ac.uk}
\affiliation{Physics, King's College London, Strand, London WC2R 2LS, UK}
\date{5th April, 2010}%
\begin{abstract}
We examine whether the accretion of dark matter onto neutron stars could ever have any visible external effects.  Captured dark matter which subsequently annihilates will heat the neutron stars, although it seems the effect will be too small to heat close neutron stars at an observable rate whilst those at the galactic centre are obscured by dust.  Non-annihilating dark matter would accumulate at the centre of the neutron star.  In a very dense region of dark matter such as that which may be found at the centre of the galaxy, a neutron star might accrete enough to cause it to collapse within a period of time less than the age of the Universe.  We calculate what value of the stable dark matter-nucleon cross section would cause this to occur for a large range of masses.
\end{abstract}
\maketitle

\section{Introduction}

Observations of the kinematics of self gravitating objects such as
galaxies and clusters of galaxies consistently send us the same message
- if we are to believe in Einstein's theory of gravity on these scales,
then there appears to be an invisible quantity of dark matter in each
of these objects which weighs more than the matter we can observe. Cosmological observations
add weight to this hypothesis and tells us that this invisible matter
cannot consist of baryons, rather it must be a new kind of matter
which interacts with the rest of the standard model rather feebly - dark matter
\cite{bergstrom}.

The exact coupling and mass of this dark matter is not known but has
been constrained. One hypothesis is that the dark matter annihilates
with itself and interacts with the rest of the standard model via
the weak interaction. This weakly interacting massive particle (WIMP)
scenario has gained favour because such particles would fall out of
equilibrium with the rest of the plasma at such a temperature that
their relic abundance today would be approximately correct to explain
the astronomical observations.

Such a scenario also predicts a direct detection signal due to the
recoil of atoms which are hit by dark matter particles, recoils which
are being searched for at several purpose built experiments (e.g.
\cite{xenon,cdms,zeplin}). We also expect to see signals from
the self annihilation of WIMP dark matter in regions of the galaxy where
the density is large, although there are many uncertainties with regards
to the magnitude of this signal.  Neither of these
signals has yet been detected although international efforts to
find such signals are intensifying to coincide with the opening of
the LHC which also may create WIMP dark matter particles.

Since we only understand the thermal history of the Universe back
to the start of nucleosynthesis, we cannot say with any surety whether
or not the WIMP scenario makes sense. Furthermore there are many other
scenarios of dark matter which involve much more massive particles
or particles which cannot annihilate with themselves \cite{wimpzillas,baryoniccharge}.  There is roughly 5-7 times the amount of dark matter in the Universe by mass relative to baryonic matter.  This ratio is rather close to one, a mystery which is only solved within the WIMP framework by a happy coincidence.  The closeness of these numbers has led some researchers to suggest that, like baryons, dark matter also possesses a conserved charge and there is an asymmetry in this charge in the Universe.  If the two asymmetries are related then one would require the dark matter mass to be approximately 5-7 times the mass of a nucleon. This intriguing possibility would be consistent with the controversial DAMA experiment \cite{dama} and the slight hint of anomalous noise in the cogent experiment\cite{cogent}.  Such a dark matter candidate could also have interesting implications for solar physics \cite{frandsen}.

Since any constraints on the nature of the mass and cross section of dark
matter particles are interesting, in this paper we will consider both of these paradigms and see whether or not it is possible to obtain any new constraints from a new angle - namely by considering the capture of dark matter by neutron stars.

The accretion of dark matter onto stellar objects has been considered
by various groups looking at both stars \cite{bouquet,freese,malcolmstars,iocco,taoso}
and compact objects \cite{moskalenkowai,mccullough,Kouvaris2007}. In particular, the ultimate fate of neutron stars which accrete
non-annihilating dark matter has been discussed before \cite{fairbairnbertone,nussinov}.

Our aim is to consider the accretion of dark matter onto neutron stars
in greater detail in order to examine whether or not it would ever
be possible to either observe the heating of a neutron star due to dark matter annihilation within the object, or the collapse of a neutron
star which accretes non annihilating dark matter.

In the next section we will outline our estimate for the accretion rate of dark matter onto a neutron star.  Then we will explain which densities we will be assuming for dark matter in the Milky Way.  We will then go on to work out how hot we can expect a neutron star to get simply due to the accretion of dark matter and compare this with observations.

Finally we will look at whether it is at all sensible to imagine a situation where the accretion of non-annihilating dark matter onto a neutron star would give rise to its subsequent collapse before concluding.

\section{Capture rate of dark matter onto Neutron stars}

In this section we will calculate the rate at which DM particles will accrete
onto neutron stars. The total capture rate depends on the density of target nuclei and the escape velocity, quantities which both vary
throughout the star. The expression which needs to be calculated
is the following \cite{gould}:
\begin{equation}
C=4\pi\int_{0}^{R_{\star}}r^{2}\frac{dC(r)}{dV}dr.
\end{equation}
 where the capture rate for a given radius is given by 
\begin{eqnarray}
\frac{dC(r)}{dV}= & \left(\frac{6}{\pi}\right)^{1/2}\sigma_{0}A_{n}^{4}\frac{\rho_{\star}}{M_{n}}\frac{\rho_{\chi}}{m_{\chi}}\frac{v^{2}(r)}{\bar{v}^{2}}\frac{\bar{v}}{2\eta A^{2}}\nonumber\\
 & \times\left\{ \left(A_{+}A_{-}-\frac{1}{2}\right)\left[\chi(-\eta,\eta)-\chi(A_{-},A_{+})\right]\right.\nonumber\\
 & \left.+\frac{1}{2}A_{+}e^{-A_{-}^{2}}-\frac{1}{2}A_{-}e^{-A_{+}^{2}}-\eta e^{-\eta^{2}}\right\} 
\label{capture}
\end{eqnarray}
 and the various functions within this expression are \[
A^{2}=\frac{3v^{2}(r)\mu}{2\bar{v}^{2}\mu_{-}^{2}},\; A_{\pm}=A\pm\eta,\;\eta=\sqrt{\frac{3v_{\star}^{2}}{2\bar{v}^{2}},}\]
 \[
\chi(a,b)=\int_{a}^{b}e^{-y^{2}}dy=\frac{\sqrt{\pi}}{2}\left[\text{erf}(b)-\text{erf}(a)\right],\]
 \[
\mu=m_{\chi}/M_{n},\;\mu_{-}=(\mu-1)/2.\]
 In the above, $m_{\chi}$ is the mass of the DM particle, $\rho_{\chi}$
is the ambient DM mass density, $A_{n}$ is the atomic number of the
neutron star nuclei, $M_{n}$ is the nucleus mass, $\bar{v}$ is the
DM velocity dispersion, $v_{\star}$ is the star's velocity with respect
to the zero point of the DM velocity distribution and $v(r)$ is the
escape velocity at a given radius $r$ inside the neutron star (see
below); subscript $\star$ refers to the neutron star quantities.

For neutron stars it is important to note that there is a maximum effective cross section -  the sum of all the cross sections of all the nuclei in the object cannot exceed the total surface area of the star since this is obviously an absolute upper limit on the total cross section of the object.  Because of this, cross sections in excess of $\sigma_{0}^{\text{max}}=2\times10^{-45}$ will not increase the probability of capture.  We will take this into account in what follows.

The quantities in this equation (\ref{capture}) have various origins - $\bar{v}$ and $\rho_{\chi}$ depend upon the distribution of dark matter in the galaxy and we will discuss them later.  The calculation of the escape velocity is more complicated -  in a normal star we can simply use Newtonian gravity but if we were to apply the same simple equations to a neutron star we would obtain superluminal escape velocities suggesting the star is unstable.  This is of course due to the fact that neutron stars are relativistic objects and we need to take into account General Relativistic effects to calculate the escape velocity properly.

The space-time geometry inside a static, spherical fluid star is
\begin{equation}
ds^{2}=-e^{2\Phi}dt^{2}+\left(1-\frac{2M(r)}{r}\right)^{-1}dr^{2}+r^{2}d\Omega^{2},\label{eq:NS_Metric}\end{equation}
 where $M(r)=4\pi\int_{0}^{r}\rho(r')r'^{2}dr'$ and $\Phi$ is determined
by solutions of
\begin{equation}
\frac{d\Phi}{dr} =\frac{GM(r)/c^{2}+4\pi Gr^{3}P(r)/c^{4}}{r\left[r-2\frac{GM(r)}{c^{2}}\right]}
\end{equation}
where $P$ is the pressure inside the star.   Following the usual Lagrange method of calculating the escape velocity we find that
\begin{equation}
v_{esc}^{2}=\frac{1}{\left(1-\frac{2GM(r)}{rc^{2}}\right)\left(\frac{d\tau}{dr}\right)^{2}+\frac{1}{c^{2}}}=c^{2}\left(1-e^{2\Phi}\right).
\label{phi}
\end{equation}
What remains is to obtain the density and pressure as a function of radius so that we are able to solve equations (\ref{capture}) and (\ref{phi}).  This is done using the Oppenheimer-Volkoff equations, which are the General relativistic versions of the equations of stellar structure.
\begin{eqnarray}
\frac{dP}{dr}&=&-\frac{G}{r^{2}}\left[\rho(r)+\frac{P(r)}{c^{2}}\right]\left[M(r)+4\pi r^{3}\frac{P(r)}{c^{2}}\right]\nonumber\\
 && \times \left[1-\frac{2GM(r)}{c^{2}r}\right]^{-1},\\
\frac{dM}{dr} & =& 4\pi r^{2}\rho(r).
\end{eqnarray}
We integrate these equations outwards for different central pressures.  The initial condition for $\Phi(r=0)$ is
chosen so that its value at the surface of the star matches its solution at
large radii $\Phi(r>R_{\text{NS}})=\frac{1}{2}\text{ln}(1-2GM/rc^{2})$.
We then vary the central pressure in order to find the maximal mass
one can obtain in the given conditions. \\
 As an equation of state (EOS) for the neutron star matter, we consider
the unified model developed by Pandharipande \& Ravenhall \cite{haensel} which is based on the Friedman-Pandharipande-Skyrme (FPS) EOS.  We then simply pick a typical solution which gives a profile apparently shared by the majority of neutron stars \cite{clarke}: $M_{\text{NS}}=1.44\, M_{\odot}$, $R_{\text{NS}}=10.6$ km,
$\rho_{\text{NS}}^{\text{central}}=1.4\times10^{18}$ kg/m$^{3}$.\\
 Once the structure of the neutron star including
the escape velocity has been obtained in this way we are half way to being able to calculate the capture rate by integrating equation (\ref{capture}).  In principle this is not a full calculation because while we are calculating the escape velocity as a function of radius in a way which makes sense, we are not then looking at the effect that curved geodesics will have on the capture rate.  Nevertheless, we believe that the capture rate calculated with the machinery presented here will be accurate within a factor of a few.   This is an appropriate level of accuracy for this work.  

We also need to make some assumption about the expected density of dark matter in the galaxy, which is what we shall turn to now.

\section{Galactic Density of Dark Matter\label{dens}}
\begin{figure}[h]
\includegraphics[scale=0.43,angle=-90]{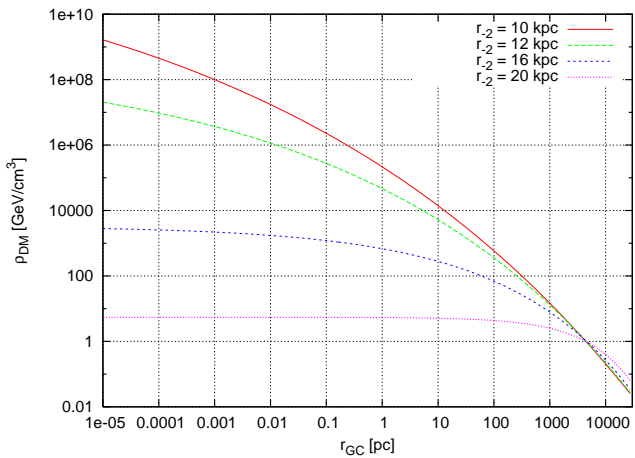}

\caption{Einasto DM density profiles for three dif\-ferent sets of parameters:
($r_{-2}$, $\alpha$) = (10 kpc, 0.06), (12 kpc, 0.09), (16 kpc,
0.19) and (20 kpc, 0.53).\label{fig:Einasto-DM-density}}

\end{figure}
There are various possible methods of obtaining a realistic profile for the density of dark matter in the galaxy.  First, we choose
to use the Einasto profile 
\begin{equation}
\rho(r)=\rho_{-2}e^{-\frac{2}{\alpha}\left[\left(\frac{r}{r_{-2}}\right)^{\alpha}-1\right]},\end{equation}
 where $\rho_{-2}$ is the DM density at galactic radius $r_{-2}$
where the logarithmic gradient $d\text{ln}\rho/d\text{ln}r=-2$ and
$\alpha$ is a parameter describing the degree of curvature of the
profile. We choose this profile because it describes well DM halos of various
sizes which appear in N-body simulations \cite{Merritt2005}.

Several recent studies have estimated the local density of dark matter in some detail \cite{ullio,deboer,salucci}.  We do not go to such lengths and adopt a simpler method, simply ensuring that the enclosed mass at the location of the sun $M_{\text{MW}}(R_{\odot})=\int_{0}^{R_{\odot}}4\pi r^{2}\rho(r)dr$
yields the correct Keplerian velocity $v_{\text{kepl.}}(R_{\odot})=\left(GM_{\text{MW}}(R)/R_{\odot}\right)^{1/2}=220$km s$^{-1}$,
where $R_{\odot}$ is the galactic radius of the Sun.  Although it neglects baryons, this gives a reasonable density in line with other methods and gives us a density of 0.3-0.5 GeV cm$^{-3}$ of dark matter at the solar radius.  One then obtains
a 1-parameter set of solutions in $\alpha$ with the corresponding
values of $r_{-2}$, which lie between 10 and 30 kpc (see
Fig. \ref{fig:Einasto-DM-density}).  Note that N-body simulations typically yield values of $\alpha\sim 0.15-0.19$ for Milky Way size halos \cite{duffy}.  We will allow a bit more freedom to adopt steeper profiles since we have not explicitly taken into account the effect of baryonic contraction in this work \cite{gnedin,jesper}.

We will assume that the velocity dispersion of dark matter is 200 km s$^{-1}$.  Sometimes it will be smaller and sometimes larger than this depending on the detailed dynamics of the dark matter halo.  However, since this is a poorly understood subject \cite{mightwe}, we will not attempt to model this in any more detail.

\section{Annihilating dark matter and its Consequences\label{sec:3.-DM-Annihilation}}
\begin{figure}[h]
\includegraphics[scale=0.43,angle=-90]{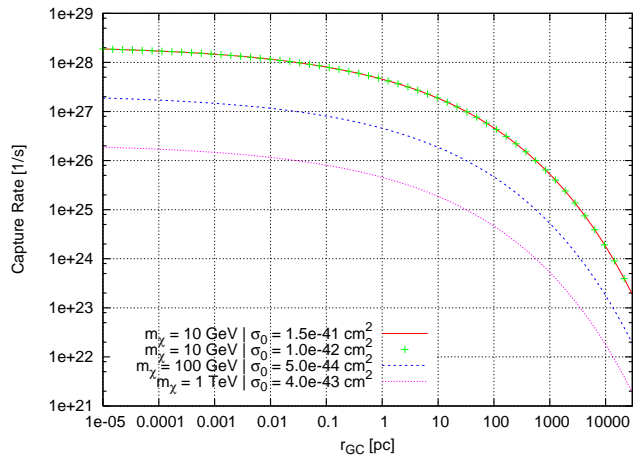}

\caption{Dark matter accretion rates vs. galactic radius for four different configurations
of dark matter (cf. legend in figure) and with ($r_{-2}$, $\alpha$) = (16
kpc, 0.19). The merging of the two first cases is due to the fact
that we take into account the limiting geometrical cross section of
the neutron star surface. \label{fig:Accretion-rates}}
\end{figure}

Now we have the neutron star density and escape velocity profiles and models for the density of dark matter in the Milky Way, we can calculate how much dark matter will be captured using equation (\ref{capture}).  The result as a function of radius for the four different density profiles can be seen in figure \ref{fig:Accretion-rates}

\subsection{Annihilation of dark matter and injection of energy}

  We next need to analyse what happens to dark matter once it has been captured.  In order to do this, we base our calculations on Kouvaris' work \cite{Kouvaris2007}, adapting
the formulae to our situation when necessary.
 For most cases of interest where the scattering cross-sections are not microscopically small, one can show that the captured WIMPs will thermalise relatively quickly,
forming a roughly Maxwell-Boltzmann distribution in velocity and distance 
around the centre of the neutron star.  If the particle we consider is an ordinary WIMP which can annihilate with itself it will do so at a rate determined by the self annihilation cross section and the density of dark matter in the star.  Following Kouvaris' argument, we assume that in most interesting situations (where the self annihilation cross section corresponds to that required for a good relic abundance), the annihilation rate reaches the accretion rate (Fig.
\ref{fig:Accretion-rates}) within around 10 million years, which is also approximately
the time where the DM annihilation begins to affect the temperature
of the neutron star.

\subsection{Cooling of the Neutron Star\protect \protect \\
 }

While some energy will be lost to neutrinos, the energy generated by the annihilation of most dark matter candidates is carried in the main by leptons, quarks and photons. We therefore neglect that part lost to neutrinos and assume that all the annihilation mass energy goes into heating the neutron star.

Since all the dominant cooling processes involved
in this situation scale with some positive power of the temperature except for the WIMP
annihilation emissivity, the latter will dominate at some point in
the life of the star.

The cooling of the neutron star is given by the following differential
equation:\begin{equation}
\frac{dT}{dt}=\frac{-\epsilon_{\nu}-\epsilon_{\gamma}+\epsilon_{\text{DM}}}{c_{V}},\label{eq:dTdt}\end{equation}
 where $c_{V}$ is the heat capacity of the star. $\epsilon_{\text{DM}}$
is the emissivity (released energy per volume per time) produced by
the annihilation of the DM when the latter saturates and is given
by $\epsilon_{\text{DM}}=3Cm_{\chi}/4\pi R^{3}$. $\epsilon_{\nu}$
is the emissivity due to the modified Urca process, which makes the
neutron star lose energy through neutrino emission by converting protons
and electrons to neutrons and vice versa, and is given by \cite{1983Shapiro}
\begin{equation}
\epsilon_{\nu}=(1.2\times10^{3}\text{ J}\,\text{m}^{-3}\text{s}^{-1})\left(\frac{n}{n_{0}}\right)^{2/3}\left(\frac{T}{10^{7}\text{K}}\right)^{8}\end{equation}
 (where $n$ is the baryon density in the star and $n_{0}=0.17\text{ fm}^{-3}$
the baryon density in nuclear matter).  Finally, $\epsilon_{\gamma}$ accounts
for the effective emissivity in photons measured in energy over
volume and time and is given by\begin{equation}
\epsilon_{\gamma}=\frac{L_{\gamma}}{(4/3)\pi R^{3}}\simeq1.56\times10^{13}\left(\frac{T}{10^{8}\text{K}}\right)^{2.2}\text{J}\,\text{m}^{-3}\text{s}^{-1},\end{equation}
 where $L_{\gamma}$ simply is the rate of heat loss from the surface
of the neutron star: $L_{\gamma}=4\pi R^{2}\sigma T_{\text{surface}}^{4}$
(with $\sigma$ the Stefan-Boltzmann constant).
\begin{figure}[h]
\includegraphics[scale=0.43,angle=-90]{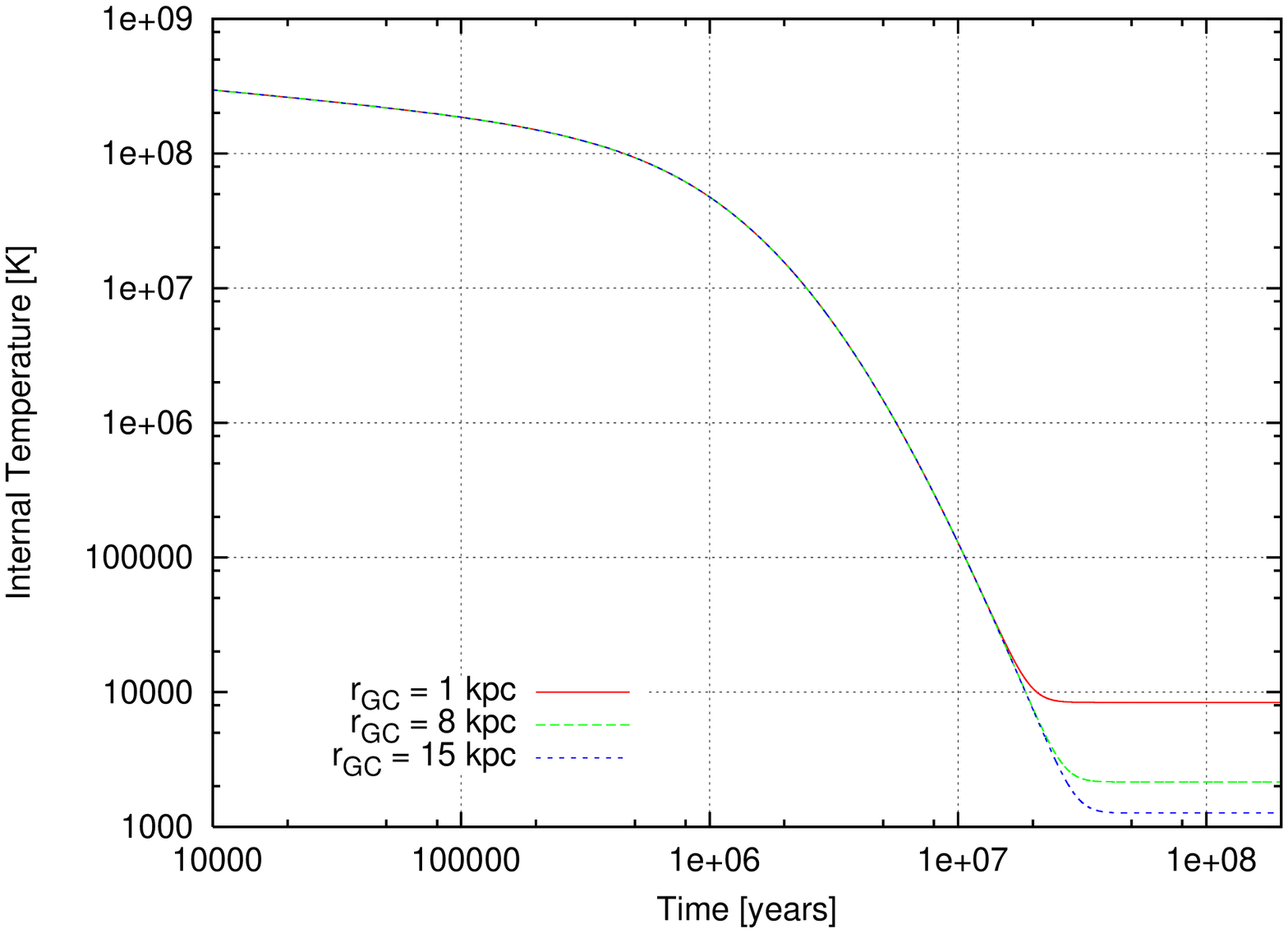}

\includegraphics[angle=-90,scale=0.43]{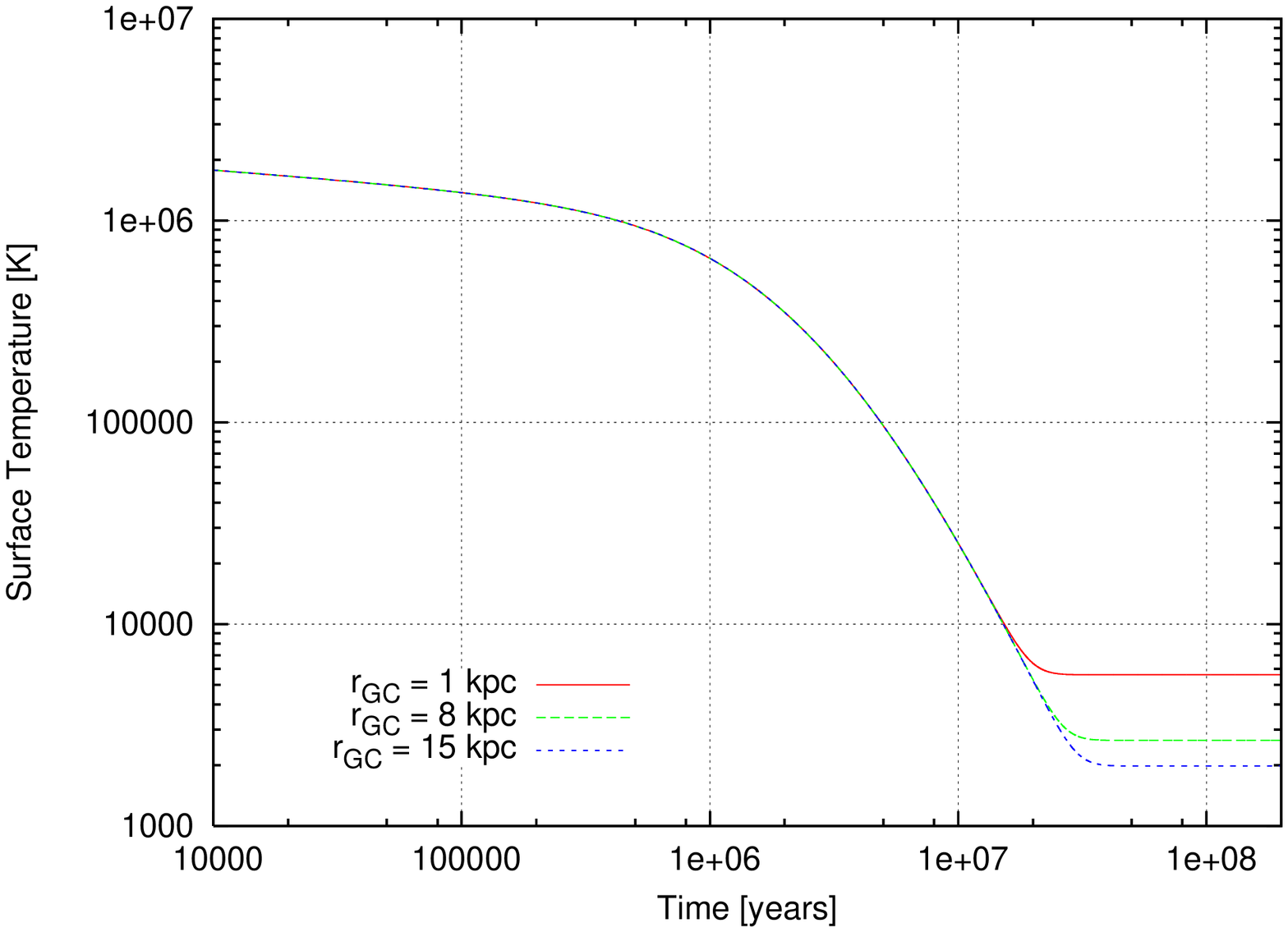}

\caption{Evolution of the internal (above) and surface temperatures of a 1.44
$M_{\odot}$ neutron star situated at various galactic radii. In the
present case, $m{}_{\chi}$= 10 GeV and $\sigma_{0}$ = $1.5\times10^{-41}$
cm$^{2}$.\label{fig:Temperatures}}

\end{figure}

\subsection{Change in outward appearance}

The DM heats the neutron star at a constant rate which will dominate any other thermal process at late times. Note that,
since when the DM particles annihilate their equilibrium density corresponds to a negligible fraction of the total mass, we have
neglected the contribution of the WIMPs to the specific heat.

In order to determine the thermal evolution of the star through its
internal and surface temperatures, we solve numerically the
differential equation (\ref{eq:dTdt}). As the temperature at interesting
times (i.e. beyond $10^{6}$ years when the star starts to cool down) is reasonably insensitive to the
initial conditions, we arbitrarily set $T(t=0)=10^{9}$ K.

The surface temperature is related to the internal temperature by
the following approximation \cite{Page2004}:\[
T_{\text{surface}}=(0.87\times10^{6}\,\text{K})\left(\frac{g_{s}}{10^{12}\,\text{m/s}^{2}}\right)^{1/4}\left(\frac{T}{10^{8}\,\text{K}}\right)^{0.55},\]
 where $g_{s}=GM_{\text{NS}}/R_{\text{NS}}^{2}$ is the surface gravity
on the neutron star.

\subsection{Results for annihilating dark matter - how hot do neutron stars get?}

Changing the position of the neutron star in the galaxy changes the accretion rate of dark matter which in turn changes the temperature of the neutron star due to internal dark matter annihilation.

Since we are working with a 1-parameter set of density profiles (see Section
\ref{dens}), our predictions will be model-dependent. However, knowing the
discrepancy between the different DM density distributions gives
us a relatively good picture of the whole space of possibilities and since many
relations involved in the process are linearly related to the
density parameter, the reader can scale our results up or down as required.

For every pair of parameters $(r_{-2},\alpha)$ defining an Einasto
profile which matches our given constraints, we define the properties
($m_{\text{DM}}$,$\sigma_{0}$) of the DM particle we want to probe;
we then calculate the capture rate for our standard neutron star and
we finally apply the temperature gradient given by Eq. (\ref{eq:dTdt})
in order to find the final internal and surface temperatures. In Fig.
(\ref{fig:Temperatures}), we present an example of our results, with
$m{}_{\chi}=10$ GeV and $\sigma_{0}=1.5\times10^{-41}$ cm$^{2}$.

Using the method described before, we calculate the final temperatures
of our neutron star at dif\-ferent galactic radii, starting with
$r_{\text{GC}}=10^{-2}$ pc and going outwards through the halo up to a radius of
50 kpc. For every DM density profile and for every type of DM particle,
we get a specific curve. As we said, however, the relation between
the various sets of solutions is rather straightforward, since the
temperature gradient varies as: $dT/dt\propto Cm_{\chi}$.

\begin{figure}[h]
\includegraphics[scale=0.43,angle=-90]{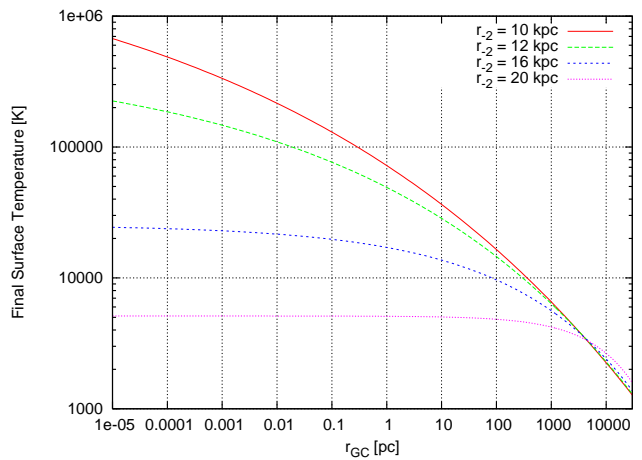}

\caption{Final surface temperature of the neutron star (at late times) for
dif\-ferent DM density distributions, with $m_{\chi}=10$ GeV and
$\sigma_{0}=1.5\times10^{-41}$ cm$^{2}$.\label{fig:Final-surface-temperatures}}

\end{figure}

In the most favorable cases in terms of DM properties and distribution,
the highest surface temperatures one can possibly obtain lie around
10$^{6}$ K near the galactic centre (Fig. \ref{fig:Final-surface-temperatures}).
Given the nature of the neutron stars, these final temperatures could
result in luminosities of the order of 10$^{-2}\, L_{\odot}$.

\subsection{Density profiles with central spikes}

In all our calculations so far we have been considering straightforward
Einasto profiles without taking too much care about the complex astrophysical
phenomena occurring near the galactic centre on a sub-parsec scale.
The formation of a super massive black hole is though to enhance the density 
in this central region \cite{gondolosilk} and this combined with effects such as self-annihilation, gravitational scattering of DM particles by
stars and capture in the supermassive black hole must be
included if one wishes to obtain a more realistic picture of this
central region.

\begin{figure}[h]
\includegraphics[scale=0.43,angle=-90]{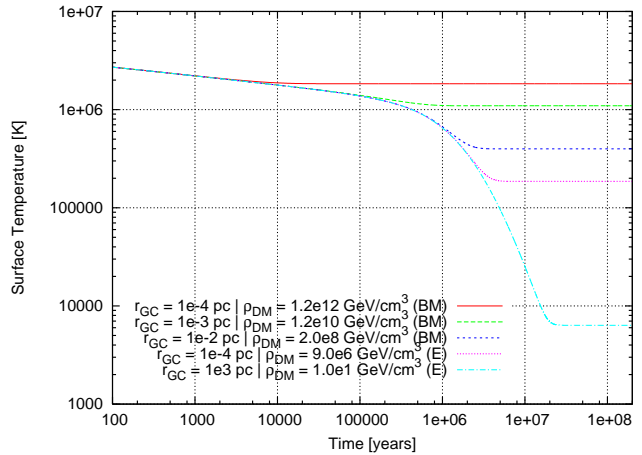}

\caption{Evolution of the surface temperature of a 1.44 $M_{\odot}$ neutron
star situated at various galactic radii (cf. legend in figure). The plot is 
for dark matter with a mass $m_\chi=10$ GeV and a scattering cross section of $\sigma_{0}$ = $1.5\times10^{-41}$.  The
DM densities are deduced from two models: profiles obtained by Bertone
\& Merritt in \cite{Bertone2005} (BM) and Einasto profiles (E).\label{fig:BM-vs-Einasto}}

\end{figure}

It is interesting to evaluate the possible consequences of considering
much higher DM densities at small radii. To do so, we take as a benchmark
the density profiles presented in \cite{Bertone2005} and we extract
a few values in order the determine the evolution of the surface temperature
(Fig. \ref{fig:BM-vs-Einasto}). One can observe that, in the most
extreme situations, the neutron star manages to keep its surface temperature
well above 10$^{6}$ K at late times if it is situated at less than
one thousandth of a parsec from the central black hole.  Since there have been a large number of compact objects observed by the Chandra X-ray telescope in this central region \cite{chandra}, this might prove interesting in the future.

\subsection{Observational Situation}
It is a challenge for astronomers to observe the thermal emission from neutron stars due to their very small surface area and luminosity.  Rather than attempting to provide a complete review of this complicated subject we will mention a couple of examples.

Pulsar PSR J0108-1431 is a nearby pulsar located between 100 and 200 pc from the Solar System.  The spin down age suggests that this object is older than $10^8$ yr so that its surface will have had a long time to cool.  Indeed, observations suggest that the surface temperature is a very low $T_{eff}<9\times 10^4$ K if the distance is 200 pc or $T_{eff}<5\times 10^4$ K if the pulsar is located at a distance of 130 pc \cite{kargaltsev}.  Comparison with figure \ref{fig:BM-vs-Einasto} shows that even this very low temperature is too high to betray underlying heating by dark matter, probably by around an order of magnitude.

Pulsar PSR J0437-4715 is a much older object with a spin down age of nearly 5 Gyr, yet its surface temperature is slightly higher than that of J0108-1431 \cite{kargaltsev}.  It seems to be challenging to explain this relatively high surface temperature (for its age) and it has been suggested that the explanation may lie in internal heating \cite{internal}.  Unfortunately, it does not seem that this internal heating comes from dark matter since in order for this to be the case we would require a density of dark matter much larger than what is expected in this region of the galaxy (PSR J0437-4715 lies only about 130 pc from the Solar System).  Also, if we were to explain the temperature of this object using dark matter, we would also expect a similar temperature for PSR J0108-1431.

If it were possible to extend observations in the future to lower temperatures and we were able to find a neutron star with a temperature $\sim 10^4$ K or lower then the situation would become much more interesting.  

As we have shown, we can expect higher temperatures for neutron stars which lie at the centre of the galaxy because of the larger accretion rates in that region.  It is however difficult to see into the centre of the galaxy, in fact the very centre is obscured by dust \cite{dust} and as we move upwards in frequency through the electromagnetic spectrum, we find the centre of the galaxy becomes invisible in the infra-red, only to return into view in the x-ray part of the spectrum.  Since this rules out the possibility of viewing objects with temperatures less than many millions of degrees, the observational situation becomes more complicated (not to mention the three orders of magnitude drop in luminosity relative to local neutron stars simply due to distance).  We may be able to see a very hot neutron star lying at the galactic centre, but it would be necessary to find some reliable way of dating the object to a relatively high precision in order to establish that this heating is anomalous and caused by dark matter.

From these considerations, it is clear that the conclusion largely depends on the assumptions made about the DM
distribution in the Milky Way. Nevertheless, there is a small hope that
the most extreme configurations might be tested
through observations.

Therefore, our chances of using neutron stars as dark matter probes
through the method described above are non-zero but limited. There exists
however an alternative way of benefiting from the high accretion power
of these very dense objects if we assume that DM particles do not
co-annihilate with each other.

\section{Non-Annihilating Dark Matter}

In the case of non-annihilating DM particles, there would be no heating of the neutron star due to annihilation - the neutron star would simply accrete DM in its core.  If the amount of dark matter in the core were to increase without limit, there would be various possible outcomes.  In the case of fermionic dark matter and in the absence of any pressure due to the exchange of gauge bosons, the density would increase until Fermi statistics starts to play a role.  When this occurs, the dark matter core would develop the equation of state of non-relativistic degenerate fermionic matter, $P \propto \rho^{5/3}$.  In order to calculate the effect of such a degenerate dark matter core upon the neutron star, the correct thing to do would be to simultaneously solve the Oppenheimer-Volkoff equation for the two stars co-located on top of each other and therefore both contributing to the gravitational field relevant for the solution.  We have obtained such solutions, and have confirmed that when there is a large mass contribution due to a degenerate dark matter core existing in the star, the normal mass-radius and indeed maximum mass of the neutron star that can be obtained varies.

In practise however, the amount of dark matter which needs to be added to the core of the neutron star for this to happen is typically far in excess of the Chandrasekhar mass ($M_{\text{Ch}}$) corresponding to the degenerate dark matter star for all but the lightest dark matter candidates.  At this critical point, the degeneracy pressure of dark matter would become relativistic and the degenerate dark matter core would become unstable.  Gravitational collapse of the dark matter core would occur, creating a black hole at the centre of the neutron star, resulting in the neutron matter also becoming unstable.

The black hole thus created would swallow the neutron star entirely.  This is particularly intriguing possibility not simply for its inherent drama but also because it could in principle provide an additional explanation of the unexplained gamma ray bursts observed in the Universe other than the coalescence of a neutron star with another compact object.  It would therefore be interesting if enough dark matter could be accreted for reasonable values of the relevant parameters - namely the density of dark matter in the places where neutron stars reside, the mass of the dark matter and its cross section for scattering off nuclei.

\begin{figure}[h]
\includegraphics[scale=0.43,angle=-90]{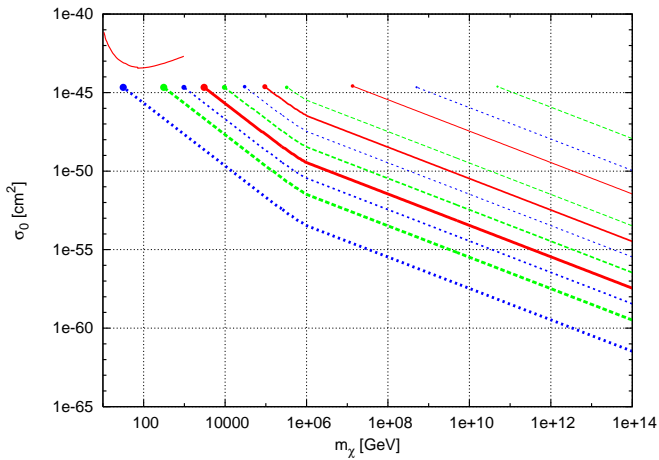}

\caption{Neutron stars immersed for long enough in a high enough density of non-annihilating dark matter will eventually accumulate
the amount of DM particles which correspond to the Chandrasekhar
limit $M_{\text{Ch}}$ and collapse.  The red solid lines here correspond to the cross section $\sigma_{0}$ and mass $m_{\chi}$ required to cause collapse after $10^6$ years.  The green longer dashed lines correspond to $10^8$ years and the blue shorter dashed lines correspond to $10^{10}$ years.  The thicker the line the higher the density - we plot four lines for each duration of time corresponding to (in decreasing thickness and therefore density) $10^{11}$, $10^{8}$, $10^{5}$ and $0.3$ GeV cm$^{-3}$.  Note the least dense red line corresponds to such a high dark matter mass it lies outside the limits of the plot to the right.  The curved line corresponds to the latest data from the CDMS experiment to place the results in context.  The change in the slope of the line in the line around $10^6$ GeV is explained in the text.
\label{fig:Sigma_0-m_DM}}
\end{figure}

Fixing the accretion time available for the neutron star to capture
DM particles, one can determine, for a given $m_{\chi}$, the cross-section
needed to accrete a mass equivalent to $M_{\text{Ch}}(m_\chi)$ which is the amount of accreted dark matter required to instigate the collapse of the Neutron star. This procedure
allows us to pick out regions on the $\sigma_{0}-m_{\chi}$ plane where this could occur (Fig.
\ref{fig:Sigma_0-m_DM}). We assume that the Chandrasekhar limit is
of the order of $M_{\text{Ch}}\approx M_{\text{Pl}}^{3}/m_{\chi}^{2}$
and we consider three accretion times: $10^{6}$, $10^{8}$ and $10^{10}$
years, corresponding roughly to 0.01\%, 1\% and 100\% of the age of the Universe respectively.  We also consider four different densities $10^{11}$, $10^{8}$, $10^{5}$ and $0.3$ GeV cm$^{-3}$.  The highest density $10^{11}$ GeV cm$^{-3}$ is the rather extreme limit of the predictions of Bertone and Merritt for non-annihilating dark matter in a central spike.
Since we consider the limiting effective cross section given by
the geometrical cross section of the neutron star ($\sigma_{0}^{\text{max}}=2\times10^{-45}$
cm$^{2}$), there exists a minimum value for the DM particle mass
below which the fixed accretion time is too short for the star to
accrete $M_{\text{Ch}}$. The corresponding limiting $\sigma_{0}$
is shown on the graph by the upper cut-off at high cross sections.
It is possible to see that in the most extreme case, namely a neutron star immersed in a density of $10^{11}$ GeV cm$^{-3}$ for $10^{10}$ years do we obtain results which are approaching the region of interest for direct detection experiments.

The cross sections required become much smaller for higher dark matter masses since much fewer of these particles need to be accreted to reach their correspondingly smaller Chandrasekhar mass limit.  The change in the slope of the line at high masses can be understood in the following way -  the typical energy exchanged in a recoil of a dark matter particle with a nucleon goes like $m_{nuc}v^2$ for $m_\chi\gg m_{nuc}$.  When a $10^3$ GeV mass dark matter particle traveling at $200$ km s$^{-1}$ falls onto a neutron star its velocity increases to close to the speed of light and the energy exchanged in this collision is much larger than the total kinetic energy the particle had at infinity.  Such a particle will lose so much energy in the kick it will almost certainly be captured by the neutron star \cite{sofia}.   For a much more massive particle this is no longer true and it will have a lower probability of being captured due to a single scatter, hence the different slope at higher energies.

\section{Conclusions}

In this work we have investigated whether or not it would ever be possible to use the accretion of dark matter onto neutron stars in order to understand the properties of dark mmatter better.  

We first looked at the effects of annihilating dark matter on the temperature of the neutron stars.  As can be seen in Fig. \ref{fig:Final-surface-temperatures}, the highest final surface temperatures which could be caused by the heating of neutron stars with dark matter lie around 10$^{6}$ K even in the most optimistic circumstances.
Given the surface area of the neutron stars, these values would produce
luminosities in the vicinity of 10$^{-2}$ $L_{\odot}$ with a peak
wavelength at about 3 nm (corresponding frequency $\sim$ 100 PHz).
These sources would thus radiate mainly in the range between extreme ultraviolet
(UV) and soft X-rays. Given the important absorption due to dust between us and the centre
of our galaxy and the presence of other luminous X-ray sources in
this region, we believe that the objects in question would prove rather
tricky to detect.

Perhaps more interesting are the constraints on non-annihilating dark matter which come from the fact that if enough of such dark matter were to accumulate onto a neutron star it would form a degenerate star at the centre.  If this internal star were to get too large, it might reach its own Chandrasekhar mass, which is smaller than that of the neutron star since the dark matter mass is greater than the nucleon mass in most models.  In the event of the mass of dark matter in the star reaching the Chandrasekhar mass of the star, the dark matter would lead to the collapse of the neutron star which collected it.  Such an event might happen for the kind of values of dark matter mass and cross section currently being probed by direct detection experiments but only in regions of extremely high density.  On the other hand, for higher mass dark matter particles, required cross sections are much smaller, since a much smaller mass of dark matter would need to be accumulated in order for collapse to occur.

The idea that the accretion of stable dark matter could be responsible for the collapse of Neutron stars is very exciting, in this paper we have quantified how likely that is.  For low mass dark matter particles, it seems extremely unlikely.

\section*{Acknowledgments}
We would like to thank Ed Cackett, Joakim Edsj\"o, Matthew McCullough, Trevor Ponman and Sofia Sivertsson for useful conversations.  As we were preparing the final version of this manuscript we became aware of a related project being carried out by Peter Tinyakov and Chris Kouvaris.  We would like to thank them for their willingness to participate in a coordinated release and we encourage the reader to compare our results with theirs.


\begin{thebibliography}{99}
\bibitem{sofia}
We would like to thank Sofia Sivertsson for helping us to understand this.
\end{thebibliography}
\end{document}